\documentstyle[12pt]{article}
\begin{document}
{\hskip 11.5cm} BIHEP-TH-94-33\\
\vspace{1ex}
{\hskip 12.0cm} AS-ITP-94-42\\
\vspace{1ex}
\vspace{8ex}
\begin{center}
{\LARGE Cyclic Family Symmetry and Lepton Hierarchy in Supersymmetry}\\
\vspace{7ex}
{\sc Dongsheng Du$^{a, b}$ ~  and ~  Chun Liu$^{a, c}$}\\
\vspace{3ex}
{\it  $^a$ CCAST (World Laboratory) P.O. Box 8730, Beijing, 100080}\\
{\it  $^b$ Institute of High Energy Physics, Academia Sinica}\\
  {\it P.O.Box  918(4),  Beijing  100039, China} $^1$\\
{\it  $^c$ Institute of Theoretical Physics, Academia Sinica}\\
  {\it P.O.Box  2735,  Beijing  100080, China
\footnote{Mailing address}}\\

\vspace{10.0ex}
{\large \bf Abstract}\\
\vspace{4ex}
\begin{minipage}{130mm}

   A cyclic symmetry among the left-handed doublets of the three families
is proposed.  This symmetry can naturally result in a realistic hierarchical
pattern of the fermion masses within the framework of supersymmetry with
nonvanishing sneutrino vacuum expectation values.\\
\end{minipage}
\end{center}

\newpage

   Within the framework of R-parity violating supersymmetry [1], in a
previous paper [2], the authors thought that, in addition to the Yukawa
interactions, the trilinear lepton number violating interactions also
contribute masses to the leptons if the sneutrinos have nonvanishing
vacuum expectation values (vevs).  This might be helpful to understand
the hierarchical pattern of the fermion masses.  However, the form of the
trilinear interactions was too arbitrary to give some definite predictions.
In this paper, we will reconsider this problem from the point of view
of symmetry among the fermion families.  And we will also discuss some
phenomenological aspects of this idea on neutrino physics.\par
\vspace{1.0cm}
   It will be interesting if there is some symmetry among the three
families.  Such symmetry can break spontaneously after the three sneutrinos
get different vevs [2], hence no exact symmetry appears explicitly in
the fermion spectra.  From the general fact that the third family of
fermions is much heavier than the other two families, the fermion mass
matrix is usually assumed to be the form of the democratic family mixing
[3] which exhibits an ${\rm S_{3L}\times S_{3R}}$ symmetry in some
basis of
eigenstates of gauge interactions.  Actually one of the ${\rm S_3}$ groups,
${\rm S_{3L}}$
or ${\rm S_{3R}}$, is enough to reflect this fact.  In the case of the
${\rm S_{3L}}$
symmetry, the fermion mass matrix has the form:
\begin{equation}
M=\left( \begin{array}{ccc}
a&b&c\\a&b&c\\a&b&c\\ \end{array} \right)
\end{equation}
Furthermore, the above matrix can be obtained under the cyclic symmetry
which is just a subgroup of ${\rm S_{3L}}$.  This order 3 cyclic group is also
called ${\rm Z_3}$.  In the following,
we will assume that there is a cyclic symmetry
among the left-handed doublets of the three families, which is denoted as
${\rm Z_{3L}}$.  Of course, this ${\rm Z_{3L}}$ symmetry
has to be slightly violated because
the mass matrix (1) is still of rank-one.  This violation can be achieved
through choosing appropriate scalar potential to make the sneutrinos have
different vevs.\par
\vspace{1.0cm}
   In this paper we still exploit the low energy supersymmetric model of
Ref. [2],
and focus on the lepton sector.  In stead of the R-parity, the baryon number
conservation is adopted.  The trilinear R-parity violating interactions
are introduced.  The supersymmetric gauge interactions can be found in text
books.  With the left-handed chiral lepton superfields and their ${\rm SU(2)
\times U(1)}$ quantum number $L_i(2,-1)$ and $E^c_i(1,2)$, the Higgs
superfields
$H_u(2,1)$, $H_d(2,-1)$ and $X(1,0)$, where $i$ denotes the family index,
the superpotential of our model which is invariant under the cyclic family
symmetry ${\rm Z_{3L}}$ is
\begin{equation}
{\cal W}=g_{Y_j}(\sum_{i}^{3}L_i^a)H_d^bE_j^c\epsilon_{ab}
+\lambda_j(L_1^aL_2^b+L_2^aL_3^b+L_3^aL_1^b)E_j^c\epsilon_{ab}
+\lambda^{\prime}X(H^a_uH^b_d\epsilon_{ab}-\mu^2)~,
\end{equation}
with $a$ and $b$ being the SU(2) indices.  In addition, the soft supersymmetric
breaking terms which include gaugino and scalar mass terms
and trilinear scalar interactions
should be added
in the Lagrangian.  It can be seen that the Yukawa couplings only depend on
the flavor of the SU(2) singlet leptons.  Such Yukawa couplings
will produce the mass matrix (1) after the Higgs fields getting nonvanishing
vevs.\par
\vspace{1.0cm}
   This ${\rm Z_{3L}}$ symmetry breaks when the sneutrino fields obtain
different vevs.  It has been shown straightforwardly that this is indeed
the case for the scalar potential with appropriate parameters [2].  All
these vevs can be taken real for our present purpose.  As for the fermion
spectra, the physical neutrinos are massless at tree level.  The mass matrix
of the charged leptons from the superpotential (2) involves the contributions
of both the Yukawa and the R-parity violating interactions:
\begin{equation}
M=\left( \begin{array}{ccc}
g_{Y_1}v_d+\lambda_1(v_2-v_3)&g_{Y_2}v_d+\lambda_2(v_2-v_3)&
g_{Y_3}v_d+\lambda_3(v_2-v_3)\\
g_{Y_1}v_d+\lambda_1(v_3-v_1)&g_{Y_2}v_d+\lambda_2(v_3-v_1)&
g_{Y_3}v_d+\lambda_3(v_3-v_1)\\
g_{Y_1}v_d+\lambda_1(v_1-v_2)&g_{Y_2}v_d+\lambda_2(v_1-v_2)&
g_{Y_3}v_d+\lambda_3(v_1-v_2)\\ \end{array} \right)~,
\end{equation}
where $v_d$ and $v_i$ are the vevs of the Higgs field and the sneutrino fields,
respectively.
As we have expected, it is the differences of the vevs of the sneutrino fields
that violate the ${\rm Z_{3L}}$ symmetry.  These differences should be much
smaller
than the vevs of the Higgs fields, this is also required by the lepton
universality.  From the mass matrix (3), we obtain the following masses for the
charged leptons:
\begin{equation}
\begin{array}{rcl}
m_{\tau} &\simeq& \sqrt{3} (g^2_{Y_1}+g^2_{Y_2}+g^2_{Y_3})^{1/2}v_d~,\\
m_{\mu}  &\simeq& a[(v_1-v_2)^2+(v_2-v_3)^2+(v_3-v_1)^2]^{1/2}~,\\
m_e      &   =  & 0~,
\end{array}
\end{equation}
where
\begin{equation}
a=[\frac{(g_{Y_1}\lambda_2-g_{Y_2}\lambda_1)^2+(g_{Y_2}\lambda_3-g_{Y_3}
\lambda_2)^2+(g_{Y_3}\lambda_1-g_{Y_1}\lambda_3)^2}
{g_{Y_1}^2+g_{Y_2}^2+g_{Y_3}^2}]^{1/2}~.\\[3mm]
\end{equation}
The tau lepton mass is determined mainly by the vev of the Higgs field $H_d$.
However at tree level, only the muon gets mass after the ${\rm Z_{3L}}$
symmetry
breaking.  The electron remains massless
because the determinant of the matrix $M$ is still zero.
By choosing the values
$v_d\sim 100$GeV, $v_i-v_j\sim (10-30)$GeV, $g_{Y_i}\sim 10^{-2}$, and
$\lambda_i \sim 10^{-2}-10^{-3}$
as a numerical illustration, the $\tau$ and $\mu$ masses can be
consistent with experimental measurements.\par
\vspace{1.0cm}
   It is interesting to find that we have naturally obtained a hierarchical
pattern for the leptons, that is $m_{\tau}\gg m_{\mu}\gg m_e$, despite the
fact that the electron is massless.  To make this pattern to be realistic,
we will consider the electron mass at the loop level.\par
\vspace{1.0cm}
   Actually this model can generate nonvanishing electron mass at the loop
level
if we further assume the soft breaking of ${\rm Z_{3L}}$.
In analyzing the neutrino magnetic moments, Refs. [4] and [5]
discussed the generation of the electron mass by supersymmetric radiative
corrections.  In a similar way, the electron mass can be induced naturally
through the one-loop diagram fig. 1 in this model, where $\chi$ and $l$
denote the neutral gauginos and the charged leptons, respectively.  The
mixing of the scalar leptons associated with different chiralities is due
to the soft breaking terms.  The structure of these mixing has the same
form as that shown in matrix (3), however, they are multiplied by a common
typical supersymmetric mass parameter $\tilde{m}$.
Fig. 1 contributes to the lepton mass matrix the following terms,
\begin{equation}
(\delta M)_{ij}=\sum_{\chi}^{}\frac{g^2_{\chi}}{16\pi^2}\frac{m_{\chi}}
{m^2_{\chi}-m^2_{\tilde{l}^c_j}}(\frac{m^2_{\chi}}{m^2_{\chi}
-m^2_{\tilde{l}_i}} \ln \frac{m^2_{\tilde{l}_i}}{m^2_{\chi}}
+\frac{m^2_{\tilde{l}^c_j}}{m^2_{\tilde{l}_i}-m^2_{\tilde{l}^c_j}}
\ln \frac{m^2_{\tilde{l}_i}}{m^2_{\tilde{l}^c_j}})\tilde{m}M_{ij}~,
\end{equation}
where $M_{ij}$ is the matrix element of Eq. (3).  The mixing of the
neutralinos have been simply neglected in writing the above equation.  In
general, the mass of the slepton $\tilde{l}_i$ receives two distinct
contributions: one from the supersymmetric couplings,
and the other from the soft breaking terms.
Both of them violate the
${\rm Z_{3L}}$ symmetry.
The determinant of the mass matrix $(M+\delta M)$ determines the electron
mass,
\begin{equation}
m_e=\frac{\det (M+\delta M)}{m_{\tau}m_{\mu}}~.
\end{equation}
With suitable ranges for the parameters, realistic value of the electron
mass can be generated, for instance, in the case of $m_{\chi}\sim 50$GeV,
$m_{\tilde{l}}\sim 100$GeV, and $\tilde{m}\sim 150$GeV.\par
\vspace{1.0cm}
   As a phenomenological application of above idea, let us turn to discuss
the neutrino masses.
In the superpotential (2), we have not introduced the term $L_iH_u$ which
can be forbidden by some other discrete symmetry.  Therefore, there is no
tree-level mixing between the neutrinos and the neutralinos.  The neutrinos
are massless at tree level in spite of the large sneutrino vevs which have
been taken in the numerical illustration.
At the one-loop level, nonvanishing Majorana neutrino
masses are produced in this model, in the same way as that given in Refs.
[4-6].  Because the ${\rm Z_{3L}}$ symmetry breaks softly in the slepton
sector,
there is no symmetry in the neutrino mass matrix, hence in general the
masses of the three neutrinos are at the same order,
\begin{equation}
m_{\nu}\sim \frac{\lambda_i^2}{16\pi^2}\frac{\tilde{m}m^2_{\tau}}
{m^2_{\tilde{l}}}~,
\end{equation}
where the sleptons have been taken to be degenerate and the squark's
contribution should have been included which we expect to be of the same
order.  With the above choices of the parameters, the neutrino mass is
typically $m_{\nu}\sim (0.1-10)$eV.  For the same reason, it seems that
this pattern of neutrino masses cannot
provide solution to the solar neutrino problem.  However, it is
cosmologically interesting that such neutrinos could be the components
of the hot dark matter [7].  It is also possible to solve the atmospheric
neutrino problem provided the lepton mixing are appropriate.  In addition,
it is hopeful to test the prediction about the neutrino mass range directly
in the double beta decay experiments in the near future.\par
\vspace{1.0cm}
   In conclusion, if the sneutrinos have nonvanishing vevs which can be
achieved through choosing appropriate scalar potential, it is possible
to understand the fermion mass hierarchy within the framework of low energy
supersymmetry.  We have proposed that there is a cyclic symmetry among the
left-handed doublets of the three families.  This symmetry results in an
interesting hierarchical pattern for the leptons, that is
$m_{\tau}\gg m_{\mu}\gg m_e$.
After considering the quantum corrections, this pattern can be realistic.
The phenomenology of this model is rich, however, it cannot be discussed
thoroughly until the quark sector is included [8].\par

\vspace{2.0cm}

   We would like to thank Z.J. Tao for many helpful discussions.

\newpage

{\Large Figure caption}\\

Fig. 1.  Supersymmetric generation of the lepton mass.

\end{document}